\documentclass[aps,prl,twocolumn,groupedaddress,amsmath,amssymb]{revtex4}

\usepackage{amssymb}
\usepackage{amsmath}
\usepackage{graphicx}
\usepackage{subfigure}
\usepackage{textcomp}
\usepackage{color}
\usepackage{amsfonts}
\usepackage{bbold}
\usepackage{dsfont}
\usepackage{epsfig}
\usepackage{hyperref}
\newcommand{\arctanh}[1]{\operatorname{arctan}}

\begin{document}

%Title of paper
%\title{Inelastic electron transport through spin coupled atomic chains}
\title{Detection of the electrostatic spin crossover effect in magnetic molecules}

\author{Aaron Hurley, Nadjib Baadji and Stefano Sanvito}
\affiliation{School of Physics and CRANN, Trinity College, Dublin 2, Ireland}

\date{\today}

\begin{abstract}
Scanning tunneling microscopy (STM) can be used to detect inelastic spin transitions in magnetic nano-structures comprising only 
a handful of atoms. Here we demonstrate that STM can uniquely identify the electrostatic spin crossover effect, whereby the exchange 
interaction between two magnetic centers in a magnetic molecule changes sign as a function of an external electric field. The fingerprint 
of such effect is a large drop in the differential conductance as the bias increases. Crucially in the case of a magnetic dimer the spin 
crossover transition inverts the order between the ground state and the first excited state, but does not change their symmetry. This 
means that at both sides of the conductance drop associated to the spin crossover transition there are two inelastic transition between 
the same states. The corresponding conductance steps split identically in a magnetic field and provide a unique way to identify the 
electrostatic spin crossover.
\end{abstract}

\pacs{75.47.Jn,73.40.Gk,73.20.-r}

\maketitle

%*********************************************************************
% Introduction
%*********************************************************************
%\emph{Introduction}

Probing and manipulating spins in solid state systems underpin the future development of spintronics and quantum 
information technology. In particular the interaction between conduction electrons and transition metal atoms with partially 
filled $d$-shells drives many low-temperature spin effects. Analyzing the conductance spectra of magnetic atoms adsorbed 
on a metallic host and probed by a scanning tunneling microscope (STM) unearths various distinctive features, which are 
indicative of many-body scattering. These include conductance steps associated to spin-flip transitions and zero-bias 
Kondo resonances. Such probing method, called spin-flip inelastic electron tunneling spectroscopy (SP-IETS), allows 
one to extract microscopic information about the spin interaction at the single atom level 
\cite{Hir1,Hir2,Otte1,Otte2,Fernandez-Rossier,Fransson,Lorente,Sothmann,Zitko1}.

So far SP-IETS experiments have all been well explained by assuming that the low-energy spin Hamiltonian for the atoms 
to probe is insensitive of the current or the STM voltage. Yet, recently it has been theoretically demonstrated that an 
electrostatic potential of sufficient strength can alter the exchange interaction between magnetically-coupled atoms 
\cite{Baadji,Diefenbach,Loss,Andrea}. In particular, it was shown that the magnitude of the energy Stark shift in a 
molecule containing two magnetic centers depends on whether the two spins are either parallel (high-spin) or antiparallel 
(low-spin) to each other, leading to a quadratic dependence of the exchange interaction between the two centers on the applied 
bias voltage \cite{Baadji}. Furthermore, when an electrical dipole breaks the molecule inversion symmetry, the spin crossover 
may occur at experimentally achievable electric fields. This effect paves the way for the development of quantum information and 
spintronics devices \cite{Timco}. Preliminary experimental evidence of such, so called electrostatic spin crossover effect (ESCE), 
has been already provided in breaking junctions experiments with organometallic molecules \cite{HzdZ}. These experiments 
however are complex since the molecule can access both different spin and charging states.

Here we investigate the possibility of using STM to detect the dependence of the exchange coupling on an electrical 
potential, by using our previously developed quantum mechanical approach based on the non-equilibrium Green's function 
formalism~\cite{Hurley,HurleyKondo,Hurleynew}. In particular, we calculate the bias-dependent conductance spectra of an 
ideal molecule comprising two exchange-coupled spin $1/2$ atoms. Previous attempts to investigate the ESCE have used 
a classical description of the localised spins \cite{Shukla}, where quantized excitations were neglected. In contrast our fully 
quantum mechanical approach allows us to study in details the elementary excitations of the system and identify the fingerprint 
of the ESCE. We will show that there exists a critical voltage, $V_\mathrm{C}$, where the conductance profile changes drastically 
and that this corresponds to the spin crossover between a low- and a high-spin state. Importantly, as the spin crossover transition 
reverses the order between the ground state and the first excited state but not their symmetry, at both side of $V_\mathrm{C}$ there 
is an inelastic conductance step between spin states having same total spin. These split in the same way in a magnetic field and 
provide a unique way to identify the ESCE-driven transition.

%\emph{Theoretical Model}

We consider a single-orbital tight-binding model~\cite{Hurley,HurleyKondo,Hurleynew}, describing a magnetic system coupled to an
STM tip (tip) and a substrate (sub). The scattering region containing the magnetic nanostructure consists of a magnetic dimer, where 
each atom $\lambda$ carries a quantum mechanical spin $\mathbf{S}_{\lambda}$ and it is characterized by an on-site energy 
$\varepsilon_0$. We assume that the tip and substrate can only couple to one atom at a time in the scattering region thus to broaden 
the electronic level $\varepsilon_0$ by $\Gamma_\mathrm{tip/sub}$ ($\Gamma=\Gamma_\mathrm{tip}+\Gamma_\mathrm{sub}$ is the 
total broadening). We model the spin-spin interaction between the localized spins by a nearest neighbour Heisenberg Hamiltonian with 
coupling strength $J_\mathrm{dd}$. Furthermore, we include interaction with an external magnetic field $\mathbf{B}$. The electron-spin 
interaction Hamiltonian is constructed within the $s$-$d$ model~\cite{Maria}, where the transport electrons are locally exchange-coupled 
to quantum spins through the exchange parameter $J_\mathrm{sd}$. Thus, the full Hamiltonian for the magnetic nanostructure comprises
three terms respectively describing the tight-binding electronic part (${H}_\mathrm{e}$), the spin part (${H}_\mathrm{sp}$) and the 
electron-spin interaction (${H}_\mathrm{e-sp}$):
\begin{align}
%\begin{eqnarray}
\label{eq:1}
&{H}_\mathrm{e}=\varepsilon_0\sum_{\lambda\:\alpha}c_{\lambda\alpha}^{\dagger}c_{\lambda\alpha}\:,\\
\label{eq:2}
&{H}_\mathrm{sp}=2J_\mathrm{dd}(V)\sum_{\lambda}^{N-1}\mathbf{S}_\lambda\cdot\mathbf{S}_{\lambda+1}+\sum_{\lambda}^{N}g{\mu_\mathrm{B}}\mathbf{B}\cdot\bold{S}_\lambda\:,\\
\label{eq:3}
&{H}_\mathrm{e-sp}=J_\mathrm{sd}\sum_{\lambda\:\alpha,\alpha'}(c_{\lambda\alpha}^{\dagger}[\boldsymbol{\sigma}_\lambda]_{{\alpha}{\alpha'}}c_{\lambda\alpha'})\cdot\mathbf{S}_\lambda\:.
%\end{eqnarray}
\end{align}
The electron ladder operators $c_{i\alpha}^{\dagger}/c_{i\alpha}$ create/annihilate an electron at site $i$ with spin 
$\alpha$ ($\alpha=\uparrow,\downarrow$), $\mu_\mathrm{B}$ is the Bohr magneton and $g$ the gyromagnetic ratio.
In equation (\ref{eq:3}) $\boldsymbol{\sigma}$ is a vector of Pauli matrices.

The electron transport problem is solved by using the non-equilibrium Green's function scheme with a perturbative treatment
of the electron-spin interaction, and where the perturbation parameter is $\alpha=J_\mathrm{sd}/\rho$ ($\rho$ is the 
electron density of states at the spin sites, see Refs.~\cite{Hurley,HurleyKondo,Hurleynew} for details). The diagonalisation of 
equation~(\ref{eq:2}) gives the eigenenergies, $\varepsilon_n$, and eigenstates, $|n\rangle$, of the spin sytem in the 
interacting region. Then, at the second order in the perturbation expansion, the transition rates, $W_{nl}$, between two 
eigenstates $|n\rangle$ and $|l\rangle$ induced by a non-spin-polarized current are given by
\begin{align}
\label{Tr:eq:2c}
W_{nl}=4\frac{\rho J_\mathrm{sd}^2}{\Gamma}\sum_{i,\eta,\eta'}|S^i_{nl}|^2\Gamma_{\eta}\Gamma_{\eta'}\zeta(\mu_{\eta}-\mu_{\eta'}-\Omega_{nl})\:,
\end{align}
where the spin matrix elements $S^i_{nl}=\langle n|S^i|l\rangle$ with $i=\{x,y,z\}$ are that of a single spin in the chain that 
is coupled to the tip (we drop the index $\lambda$). Furthermore $\zeta(x)=x/(1-e^{-x/k_\mathrm{B}T})$ and $\mu_{\eta}$ is 
the chemical potential in $\eta$-th lead, $\eta=\{\mathrm{tip,sub}\}$. We also assume that the onsite energy is large enough 
for the density of states of the sample to remain constant in the small energy window of interest and therefore 
$\rho=\Gamma/(\varepsilon_0^2+\Gamma^2)$. The above transition rates can be used to evaluate the bias-dependent 
non-equilibrium population, $P_n$, of the spin states, $|n\rangle$, as the steady state solution of the following equation
\begin{align}
\label{dPn:eq:1}
\frac{\mathrm{d}P_n}{\mathrm{d}t}=&\sum_{l}\Big[P_n(1-P_l)W_{ln}-P_l(1-P_n)W_{nl}\Big]\nonumber \\
&+(P_n^0-P_n)/k_\mathrm{B}T\:,
\end{align}
where $T$ is the temperature and $k_\mathrm{B}$ the Boltzmann constant. As $P_n$ enters in the definition of the Green's function 
for the transport electrons, Eq.~(\ref{dPn:eq:1}) needs to be iterated self-consistently together with that of the electron 
propagator (see Ref.~\cite{Hurley} for details).

At the second order the normalized current, $I$, flowing through a single spin can be expressed~\cite{Hurley} as a function of the 
potential bias, $V$, as
\begin{align}
\label{IV:eq:2a}
I(V)=V+\frac{\rho J_\mathrm{sd}^2}{\Gamma}\sum_{i,m,k}|S^i_{nl}|^2P_k(V)i_{-}(V-\Omega_{mk})\:,
\end{align}
where the normalisation constant is given by the elastic conductance 
$G_0=\frac{2e^2}{h}(\frac{\rho\Gamma_\mathrm{tip}\Gamma_\mathrm{sub}}{\Gamma})$
and where we have defined the quantity $i_{-}(V-\Omega_{mk})=\zeta(V-\Omega_{mk})-\zeta(-V-\Omega_{mk})$.
In addition to the elastic contribution, the total normalized conductance, $G(V)=1+G_1(V)+G_2(V)$, includes two 
parts, respectively
\begin{align}
\label{eq:2b}
&G_1(V)=\frac{\rho J_\mathrm{sd}^2}{\Gamma}\sum_{i,m,k}|S^i_{nl}|^2P_k(V)\frac{d}{dV}i_{-}(V-\Omega_{mk})\:,\\ 
\label{eq:2c}
&G_2(V)=\frac{\rho J_\mathrm{sd}^2}{\Gamma}\sum_{i,m,k}|S^i_{nl}|^2i_{-}(V-\Omega_{mk})\frac{d}{dV}P_k(V)\:.
\end{align}
At equilibrium $P_k$ does not depend on $V$ and $G_1$ becomes the only contribution to $G(V)$. This provides a conductance step 
whenever the voltage coincides with inelastic energy transition $\Omega_{mk}=|\varepsilon_m-\varepsilon_k|/e$, being $e$ the electron
charge. The conductance step height over the elastic conductance is then governed by the ratio ${\rho J_\mathrm{sd}^2}/{\Gamma}$. 
However in the non-equilibrium case $P_k(V)$ is not constant and both $G_1$ and $G_2$ contribute, resulting in more complicated 
non-linear conductance profile.

We now consider a model molecule undergoing the ESCE. This is described by a dimer of $S$=$1/2$ spins exchanged coupled through 
the bias-dependent exchange parameter $J_\mathrm{dd}(V)=J^0_\mathrm{dd}+bV^2$~\cite{Baadji}. The parameter $J^0_\mathrm{dd}$ 
is the exchange in absence of an electric field, while $b$ defines the value of the critical voltage, 
$V_\mathrm{C}=\pm\sqrt{-J^0_\mathrm{dd}/b}$, at which the magnetic coupling switches from ferromagnetic ($J_\mathrm{dd}>0$) to 
anti-ferromagnetic ($J_\mathrm{dd}<0$). Here we consider, in line with density functional theory predictions \cite{Baadji}, 
$J^0_\mathrm{dd}=6$~meV and $b=-0.04$~meV/(mV)$^2$, which give us $V_\mathrm{C}\sim$~12.25~meV. The other parameters of 
the model are fixed either from experimental data or from theoretical considerations for systems presenting a similar STM setup. 
Firstly we consider an $s$-$d$ exchange interaction $J_{\mathrm{sd}}=500$~meV, as found from theory \cite{Lucignano}. We further 
assume that the electronic energies of the dimer atoms are of the order of $1$~eV, far enough from the Fermi energy. This produces a 
constant density of states $\rho$ in the low-energy window of interest to the transport. The atoms are coupled to the underlying substrate 
strongly enough to produce a broadening of the onsite energies, $\Gamma_{\mathrm{sub}}=250$~meV. Furthermore, we assume 
that the tip is non spin-polarized and movable such that conductance spectra can be investigated for a range of $\Gamma_{\mathrm{tip}}$. 
We will also examine the effect of magnetic fields on the spectra and fix the Lande $g$-factor to 2. All spectra are obtained at a temperature 
of 1.5~K.

\begin{figure}[h]
\centering
\resizebox{\columnwidth}{!}{\includegraphics[width=7cm]{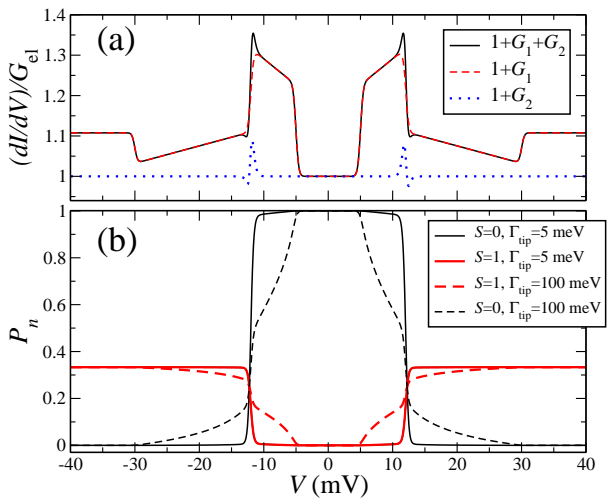}}
\caption{\footnotesize{(Color online) Conductance spectrum for a magnetic dimer exchanged coupled through a voltage-dependent
exchange parameter $J_\mathrm{dd}$. (a) total normalised conductance and the different contributions to the total conductance. The 
conductance spectrum is calculated by assuming a week tip-to-atom electronic coupling, $\Gamma_{\mathrm{tip}}=5$~meV. Note 
the large dip in the conductance at $V_\mathrm{C}\sim$~12~mV, corresponding to the spin crossover transition. (b) Population of the 
singlet (black line) and triplet (red line) states for weak, $\Gamma_{\mathrm{tip}}=5$~meV (solid line), and strong, 
$\Gamma_{\mathrm{tip}}=100$~meV (dashed line), tip-to-atom coupling.}}
\label{1}
\end{figure}

The top panel of figure~\ref{1} shows the total conductance calculated when the tip is positioned above one of the 
atoms forming the dimer, and the tip-to-atom coupling is weak, $\Gamma_{\mathrm{tip}}=5$~meV. In the same panel 
we show how the spectrum is decomposed into its various parts, $G_1$ and $G_2$ [see equations (\ref{eq:2b}) and (\ref{eq:2c})]. 
The conductance spectrum displays a first conductance step at about 5~mV, corresponding to the singlet to triplet spin-excitation. 
Then at the spin crossover critical voltage, $V_\mathrm{C}$, there is a second step in the opposite direction, i.e. the conductance 
drastically drops. This feature can be easily understood by looking at the population of the spin states as a function of bias,
also plotted in Fig.~\ref{1}, panel (b). 
%
%\begin{figure}[h]
%\centering
%\resizebox{\columnwidth}{!}{\includegraphics[width=5cm,angle=-90]{Fig2.pdf}}
%\caption{\footnotesize{(Color online) }}
%\label{2}
%\end{figure}

For $V<V_\mathrm{C}$ the molecule is in the singlet state, which is occupied with $P=1$. Note that the inelastic transition at 
$V\sim5$~mV does not alter the spin population since the current is small and not spin-polarized, so that it does not produce 
spin pumping. In other words the spin system always relaxes back to the ground state in between two inelastic scattering events.
At the crossover voltage a change in the ground state occurs and the population of the singlet state rapidly decreases to zero, 
while that of the triplet states reaches 1/3. Note that the triplet state is three times degenerate, meaning that each of the third 
component of the $S=1$ state are equally occupied (there is no magnetic anisotropy in the model). Finally, as the bias is further 
increased a third conductance step appears at about 30~meV. This is now related to an inelastic transition between the new 
triplet ground state and the first singlet excited state, i.e. between the same spin states responsible for the transition found at 
5~mV, whose energy order is now reversed. Again there is no spin pumping and the spin population remains unaffected.

From the figure it is rather clear that most of the spectral features originate from $G_1$, which does not depend on the derivative 
of the spin population. The effect of $G_2$ is evident as an additional highly non-linear contribution to the conductance at the 
spin crossover voltage, brought about by the derivative of the population of the spin states.

\begin{figure}[h]
\centering
\resizebox{\columnwidth}{!}{\includegraphics[width=5cm,angle=-90]{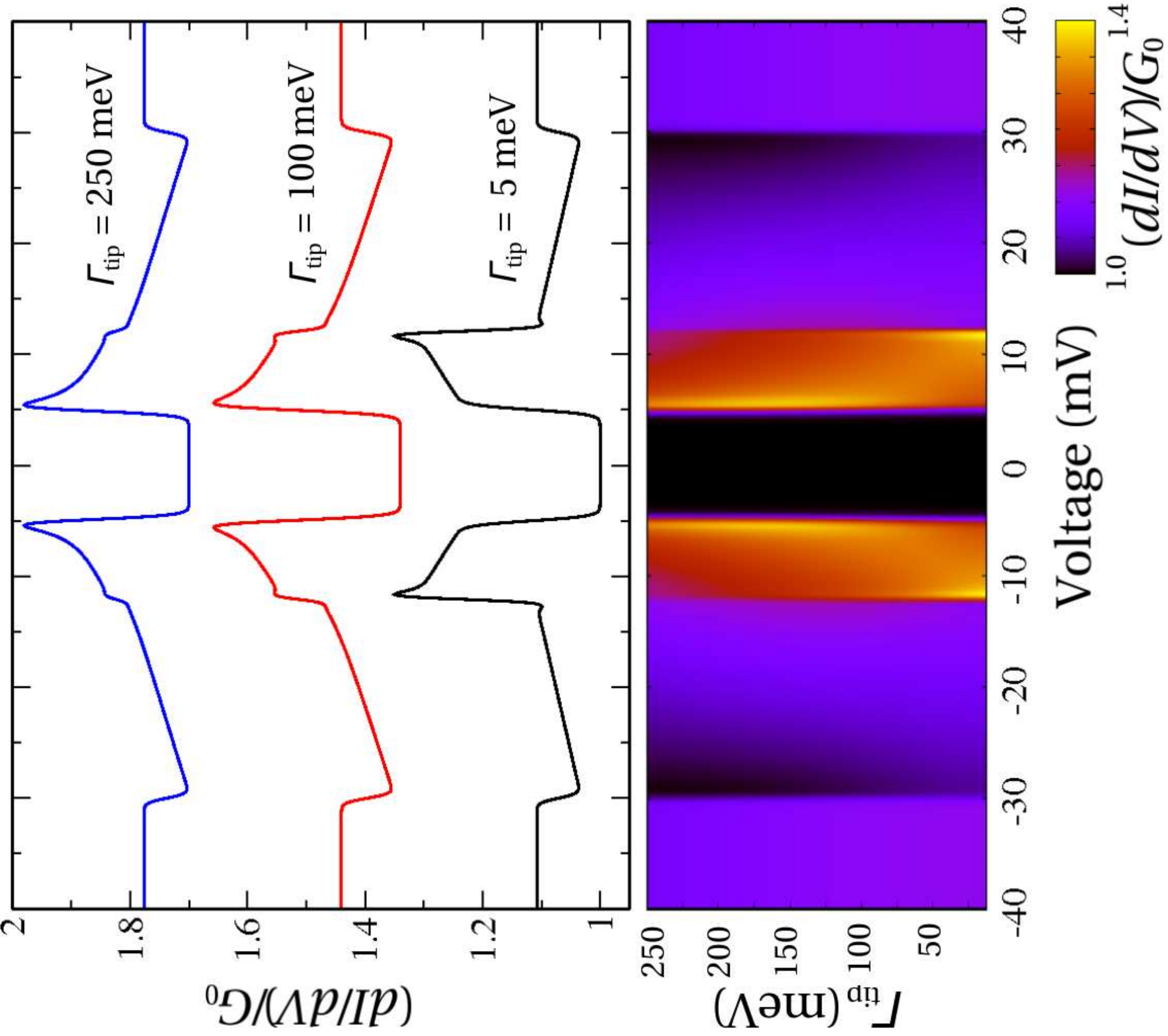}}
\caption{\footnotesize{(Color online) Conductance spectra for a spin 1/2 dimer undergoing the ESCE. In the upper panel we show 
the normalised conductance spectra for three different values of $\Gamma_\mathrm{tip}$, while the contour plot in the lower panel
show how the spectrum changes continuously between 0 and 250~mV. Note the weakening of the conductance drop at $V_C$ 
as the tip-to-atom coupling gets larger.}}
\label{2}
\end{figure}

We now investigate how the line-shape changes as the STM tip is brought into closer contact to the molecule, and Fig.~\ref{2} reports 
$G(V)$ for values of $\Gamma_\mathrm{tip}$ in the 5-250~meV range. 
From the figure one can clearly see that the conductance profile gets drastically modified, in particular in the spectral region near 
the spin crossover transition. Again the results can be understood by looking at the evolution of the spin population as a function of bias, 
plotted for $\Gamma_\mathrm{tip}=100$~meV in Fig.~\ref{1}(b). The most striking feature is that there is a significant triplet (singlet) population 
for $V<V_\mathrm{C}$ ($V<V_\mathrm{C}$), i.e. that the more intense current is now able to maintain the molecule in a steady state in 
which both the ground state and the excited state are partially occupied. It is also not surprising that the mixing is maximized at around 
$V_\mathrm{C}$ since the exchange coupling is reduced and the two spin states are less energetically separated. This feature results in 
an enhancement of the first conductance step followed by the associated conductance decay~\cite{Sothmann,Hurleynew}.

\begin{figure}[h]
\centering
\resizebox{\columnwidth}{!}{\includegraphics[width=5cm,angle=-90]{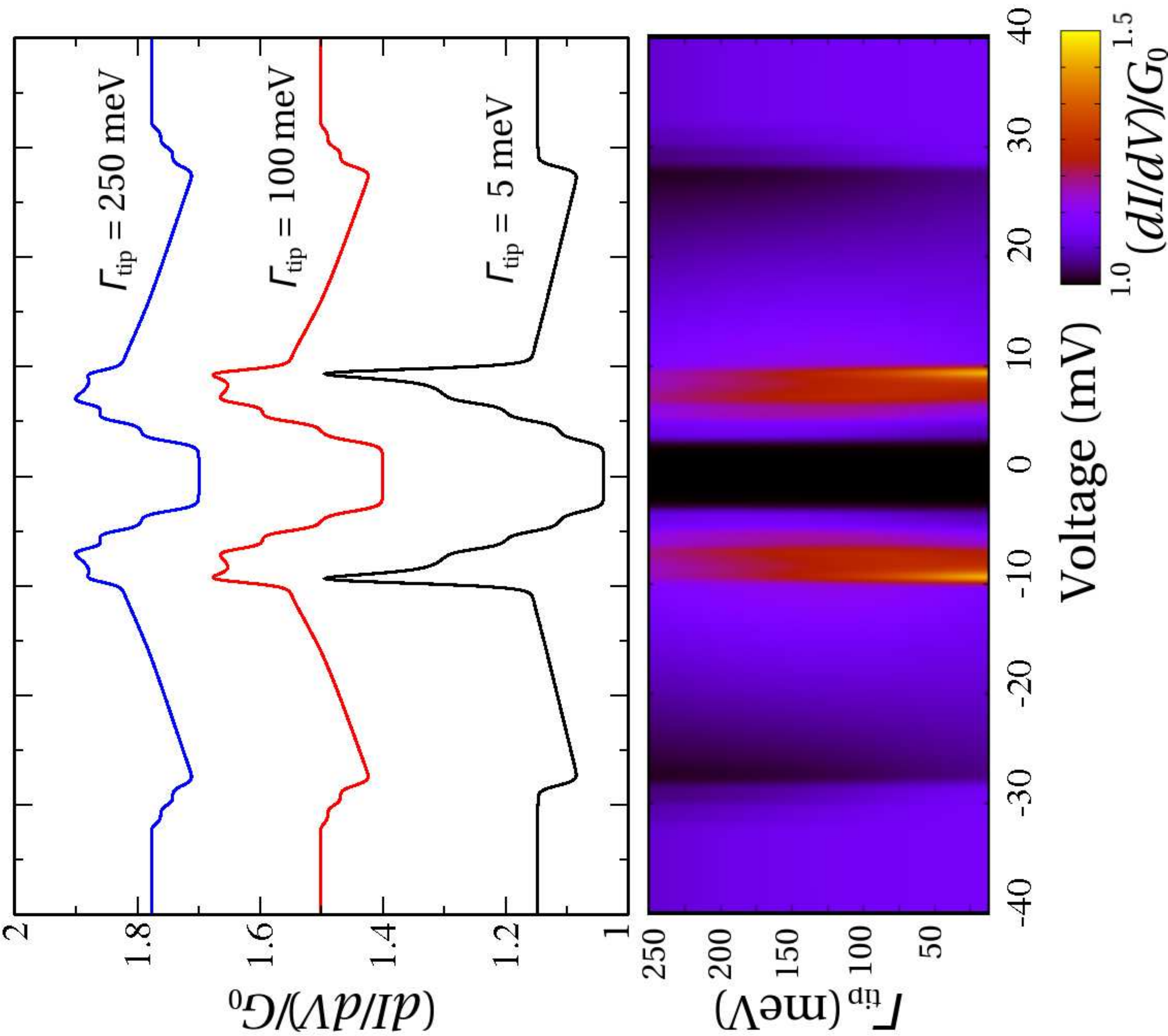}}
\caption{\footnotesize{(Color online) Conductance spectra for a spin 1/2 dimer undergoing the ESCE and subject to a magnetic field
of 20~T. In the upper panel we show the normalised conductance spectra for three different values of $\Gamma_\mathrm{tip}$, while 
the contour plot in the lower panel shows how the spectrum changes continuously between 0 and 250~mV. Note that the conductance 
steps corresponding to two inelastic transitions at both side of the spin crossover transition split into three equally spaced and 
equally intense steps. This is due to the Zeeman splitting of the triplet state.}}
\label{3}
\end{figure}
Finally we investigate the effects of applying a magnetic field and for the purpose of illustration we chose the rather large value of 
20~T (at the limit of what experimentally achievable). Fig.~\ref{3} shows the evolution of the conductance spectrum as a function of 
$\Gamma_{\mathrm{tip}}$. We notice that the onset of the magnetic field produces three equal sized and equally spaced conductance 
steps at around 5~mV. These originate from transitions between the $S=0$ ground state and the three Zeeman-split components of 
the $S=1$ excited state. Also in this case the spin crossover transition produces a drastic drop in the conductance, which becomes 
less prominent as the tip-to-atom interaction strength gets larger. Intriguingly the same splitting occurs for the second conductance 
step at $\sim$30~mV, although in this case the Zeeman-split triplet states are resolved only for large tip-to-atom interaction. Note 
that, although for the inelastic transition at 5~mV the initial state is $S=0$ and the final is $S=1$ while for the one at 30~mV the order 
is reversed, in both cases the associated conductance step splits with an identical multiplicity. This is because the ESCE simply 
inverts the order of the two spin states involved in the transition but does not change their symmetry. Thus the presence in the 
conductance spectrum of a drastic drop located in between two inelastic transitions, which split identically in a magnetic field, is the 
fingerprint of the ESCE and provides a unique way for identifying it. 

In conclusion we have investigated the possibility of detecting, by a spectroscopical STM experiment, the electrostatic spin crossover 
effect, whereby the ground state of a molecule changes between singlet to triplet as the potential voltage increases. The hallmark of 
the spin crossover transition is a drastic conductance drop, following the first conductance step
associated to the singlet to triplet transition. Importantly as the ground state changes from singlet to triplet one expects a second step
corresponding to the opposite transition (from triplet to singlet). One then has the situation in which the spin crossover conductance 
drop is always placed in between two inelastic steps both associated to transitions between the singlet and the triplet. Such transitions 
can be easily identified by applying a magnetic field as Zeeman splitting resolves the step into three equally spaced and equally sized 
steps. 

This work is sponsored by Science Foundation of Ireland (grant 08/ERA/I1759) and CRANN. Computational resources have been provided 
by the Trinity Centre for High Performance Computing (TCHPC).


\small
\begin{thebibliography}{}

\bibitem{Hir1}
  C. F. Hirjibehedin, C. P. Lutz and A. J. Heinrich, Science \textbf{312}, 1021 (2006).

\bibitem{Hir2}
C.F.~Hirjibehedin,  \textit{et al.},
%C.-Y.~Lin, A.F.~Otte, M.~Ternes, C.P.~Lutz, B.A.~Jones, and A.J.~Heinrich,
Science \textbf{317}, 1199 (2007).

\bibitem{Otte1}
  A. F. Otte, \textit{et al.}, Nature Physics \textbf{4}, 847 (2008).

\bibitem{Otte2}
  A. F. Otte,  \textit{et al.}, Phys. Rev. Lett. \textbf{103}, 107203 (2009).

\bibitem{Fernandez-Rossier}
J. Fernandez-Rossier, Phys. Rev. Lett. \textbf{102}, 256802 (2009).

\bibitem{Fransson}
J. Fransson, Phys. Rev. B \textbf{81}, 115454 (2010).

\bibitem{Lorente}
N. Lorente and J. Gauyacq, Phys. Rev. Lett. \textbf{103}, 176601 (2009). 

\bibitem{Sothmann}
B. Sothmann and J. Konig, New J. Phys. \textbf{12}, 083028 (2010).

\bibitem{Zitko1}
  R. Zitko, \textit{et al.}, New J. Phys.  {\textbf{11}}, 053003 (2009).

\bibitem{Baadji}
  N. Baadji \emph{et al.}, Nature Materials {\bf 8}, 813 (2009) .

\bibitem{Diefenbach}
 M. Diefenbach and K.S. Kim, Angew. Chem., Int. Ed. {\textbf{46}},7640 (2007).

\bibitem{Loss}M.~Trif, F.~Troiani, D.~Stepanenko and D.~Loss,
Phys. Rev. Lett. {\bf 101},  217201 (2008).

\bibitem{Andrea}A.~Droghetti and S.~Sanvito,
Phys. Rev. Lett. {\bf 107}, 047201 (2011).

\bibitem{Timco}
 G.A. Timco \textit{et al.}, Nature Nanotechnol. {\textbf{4}}, 173 (2009)

\bibitem{HzdZ}V.~Meded \textit{et al.}, 
%A.~Bagrets, K.~Fink, R.~Chandrasekar, M.~Ruben, F.~Evers, A.~Bernand-Mantel, J.S.~Seldenthuis, A.~Beukman and H.S.J.~van~der~Zant, 
Phys. Rev. B {\bf 83}, 245415 (2011).

\bibitem{Hurley}
 A. Hurley, N.Baadji and S. Sanvito, Phys. Rev. B 84, 035427 (2011)

\bibitem{HurleyKondo}
 A. Hurley, N.Baadji and S. Sanvito, Phys. Rev. B 84, 115435 (2011)

\bibitem{Hurleynew}
 A. Hurley, N.Baadji and S. Sanvito, Phys. Rev. B {\bf 86}, 125411 (2012)
 
\bibitem{Shukla}
 S. K. Shukla and S. Sanvito, Phys. Rev. B 80, 184429 (2009)

\bibitem{Maria}M.~Stamenova, T.N.~Todorov and S.~Sanvito, Phys. Rev. B {\bf 77}, 054439 (2008).

\bibitem{Lucignano}
  P. Lucignano \textit{et al.}, Nature Materials, \textbf{8}, 563 (2009).

\end{thebibliography}
\end{document}